\documentstyle[sprocl]{article} 
  \def\bea{\begin{eqnarray}}
\def\eea{\end{eqnarray}}
  \newcommand\beq{\begin{equation}}
 \newcommand\eeq{\end{equation}} \newcommand\beqn{\begin{eqnarray}}
 \newcommand\eeqn{\end{eqnarray}}

\begin{document}

\title{Highlights of the Theory\footnote{
Summary report on Working Group 5B, Theory, at DIS'98
Brussels, April 4-8, 1998}}

\author{Boris~Z.~Kopeliovich$^{1,2}$ and Robert Peschanski$^{3}$}

\vspace*{0.5cm}

\address{$^1$~Max-Planck-Institut f\"ur Kernphysik Postfach 30980, 69029
Heidelberg, Germany}

\address{$^2$~Joint Institute for Nuclear Research 141980 Moscow Region, Dubna,
Russia}

\address{$^3$ Service de Physique, CEA, CE-Saclay F-91191 Gif-sur-Yvette
France}

\maketitle\abstracts{The highlights of the Working Group 5B, {\it Theory} are
summarised. There has been reported impressive progresses in the study of BFKL
Pomeron and Odderon dynamics. It turns out that the leading log approximation
is not still a good one in the energy range of HERA and one should take care of
next to leading corrections.}

\section*{BFKL dynamics}

New results on the BFKL dynamics presented at the session WG5B {\it Theory}
initiated a hot discussion.  Viktor Fadin reported his and Lev Lipatov's
calculations for the next-to-leading correction (NLLA) to the leading-log
approximation (LLA) completed recently \cite{fl,fadin} (see also \cite{cc}).
The main result is the kernel of the BFKL equation \cite{fadin}.  Due to scale
invariance the LLA kernel has a simple complete set of the eigenfunctions
$(q_2^2)^{\frac{1}{2}+i\nu}\,$ with the largest eigenvalue $\omega_P ^B=4N_c\ln
2\alpha _s/\pi$ equal to the Pomeron intercept at the symmetric point $\nu=0$.
The argument of the running coupling is not fixed in this approximation.

The rough estimate of the influence of the (rather involved) correction to the
kernel can be made similar in a way to the calculation of the bound-energy
shift in nonrelativistic quantum mechanics, i.e.by taking the average value of
a disturbance in a non-disturbed state. This naive estimate gives for the
corrected intercept:
\begin{equation}
\omega_P \simeq \omega_P^B(1-2.4\omega_P^B).
\label{1}
\end{equation}
The coefficient 2.4 does not look large.  It fits with the rapidity interval
where correlations, neglected at LLA, become important
in
particle production processes.  This NLLA correction would be small if the LLA
intercept $\omega^B_P$ were small.  This is, unfortunately, not a case, and for
$\omega^B_P=0.4$ the two terms in brackets in (\ref{1}) nearly cancel.
It
is worth noting, however, that, firstly, this estimate is quite straightforward
and does not take into account neither the influence of the running coupling on
the eigenfunctions nor the nonperturbative effects; secondly, the value of the
correction strongly depends on its presentation.  For example, if one takes
into account the NLLA correction by the corresponding increase of the argument
of the running QCD coupling constant, $\omega_P$ at $\alpha _s(q^2)=0.15$ turns
out to be only twice smaller than its Born value.  Another possibility which
was mentioned during the discussion by Lev Lipatov is that the symmetric point
$\nu=0$ does not correspond anymore to the maximal eigenvalue of the kernel
after the NLLA correction is included.  Such a possibility is indeed realistic.
It was shown in \cite{ross} that the symmetric point corresponds to the maximum
only for very small $\alpha_s$ where the correction is also small, otherwise
the maximal eigenvalue which determines the high energy behaviour occurs at a
nonzero value of $\nu$ which depends on the coupling constant.  The
corresponding contribution to the cross section is sizable: for $\alpha_s\simeq
0.22$ it gives a Pomeron intercept $\omega_P=0.2$ consistent with experimental
data on diffractive deep-inelastic scattering.

At the moment the situation with the BFKL-based phenomenology seems to be
confusing. This difficulty was stressed by Richard Ball during the following
discussion. He presented his and Stefano Forte's results on NLLA corrections to
anomalous dimensions which turn out to be enormous. Their pessimistic
conclusion based of the failure of LLA is that there is no point to sum over
powers of $log(1/x)$ evolving the parton distributions, but one should try to
develop a new
approach.
On contrary, Al Mueller was advocating the possibility of a resummation formula
leading to a displaced but well-defined BFKL singularity.  All in all, the need
for a better conceptual understanding of the beautiful achievement of NLLA BFKL
calculations was the major outcome of the lively discussion.

If abstracting from the problem of NLLA corrections, the BFKL Pomeron in the
leading order is a sum of moving Regge poles and the most interesting is the
rightmost one.  Other poles, however, substantially affect the preasymptotic
behaviour of the proton structure function in the kinematical region of HERA. A
way to disentangle the leading pole contribution was suggested in the talk of
Volodya Zoller. In the dipole impact parameter representation the subleading
poles have nodes in dipole representation at $r_T \approx 0.5 - 1\
fm$. Therefore, the rightmost pole dominates the amplitude in this range of
transverse separations. It is suggested that the charm structure function,
which corresponds to interaction of a $c\bar c$ fluctuation of the virtual
photon, is especially sensitive to this region of $r_T$ and provides unique
information about the leading energy dependence of the BFKL Regge
poles. Vanishing contribution of the sub-leading BFKL poles at $Q^2 \leq 10\
GeV^2$ might have relevance to the deviation of $dF_2/d\, \log(Q^2)$ from the
behaviour which follows from the evolution equations, the experimentally
observed effect presented at this session by A.~Caldwell.

However, all these results are obtained for the LLA BFKL and the importance of
NLLA corrections make questionable their relevance to phenomenology.

We also note the derivation of an exact solution of the BFKL equation in {\it
three} dimensions by Dimitri Ivanov reporting calculations with Lev Lipatov and
co-workers. Similarities and differences with the four-dimensional case were
discussed.

\section*{The Odderon intercept}

A generalization of the BFKL equation for the Pomeron intercept
\begin{equation}
E\,\Phi=H_{12}\,\Phi\ ,
\label{2}
\end{equation}
where $\Phi\propto s^{\Delta}$,
$\Delta=-g^2N_cE_{min}/8\pi^2=(g^2/\pi^2)N_c\,ln2$, is the BKP
(Bartels-Kwiecinski-Praszalowicz) equation for a multi-gluon exchange,
\begin{equation}
E\,\Phi(k_1,...k_n)= \sum\limits_{i<k} H_{ik}\,\frac{T^a_iT^a_k}
{(-N_c)}\,\Phi\ ,
\label{3}
\end{equation}
where $T^a_i$ are the generators of $SU(N_c)$ group acting on the gluon color
indexes $i$.  A solution of the BKP equation could settle the problem of
unitarization of the BFKL Pomeron (a generalized ''eikonalization''). In
analogy to the conformal $SL(2,C)$ invariance which helps to solve the BFKL
equation, one should look for symmetries of the BKP equation. A
duality
symmetry
of Reggeon interactions, similar to the famous electric-magnetic duality of
gauge field theories, in was discussed by Lev Lipatov in his talk.

Being an extension to many gluon channels of the two-gluon BFKL equation, the
BKP also contains a solution with negative $C$-parity in $t$-channel, a so
called Odderon.  The minimal number of exchanged interacting gluons is three.
A solution for the BKP and for the Odderon intercept was presented by Romuald
Janik. The result turns out to be negative,
\begin{equation}
\omega_O = -0.16606\, \frac{3\alpha_sN_c}{2\pi}
\label{4}
\end{equation}
The Odderon can contribute to nucleon-nucleon elastic scattering, $K_L \to K_S$
regeneration, etc. Eq.~(\ref{4}) predicts decreasing energy dependence for such
a contribution. Since the Odderon amplitude is mostly real a good chance to
detect it is the interference with the real part of the Pomeron amplitude in
the vicinity of the minimum in the differential cross section of $NN$ elastic
scattering at $|t|\approx 1.3-1.5\ GeV^2$. Such an interference has different
signs for $pp$ and $\bar pp$ scattering and can explain the observed difference
between $ISR$ and $Sp\bar pS$ results. However, besides the Odderon intercept
one needs to know the $t$-dependence. Therefore, it seems to be problematic to
find an observable manifestation of (\ref{4}), at least for the time being.

\section*{Renormalization Group and Fracture Functions}

In the better-known and fruitful domain of the Renormalization Group properties
of QCD, Massimo Grazzini reported the progress made in the derivation of
evolution equations for the {\it Fracture Functions}.  Fracture Functions are
one-particle inclusive observables which correspond to a mixture of both
structure and fragmentation functions. They give the probability
$M(x_{Bj},x_{Fey},Q^2)$ of finding a hadron with a certain Feynman fraction
$x_{Fey}$ of outgoing momentum once there has been a deep-inelastic interaction
of a quark with Bjorken fraction $x_{Bj}$ of ingoing momentum on a virtual
photon of virtuality $Q^2.$ Besides the interest of joining two great men in the
same observable, fracture functions have the property of obeying evolution
equations dictated by perturbative QCD and the renormalization Group.

The new point put forward by Grazzini is the elucidation of a puzzle about
these evolution equations. While the original $M(x_{Bj},x_{Fey},Q^2)$ is driven
by quite non-standard evolution equations with an inhomogeneous term, it was
shown that the more differential $M(x_{Bj},x_{Fey},Q^2;t),$ where $t$ is the
momentum-squared transferred to the proton target, follows standard DGLAP
renormalization Group equations at leading order.

For the future, interesting questions remain such that higher-order corrections
or the applicability of the formalism to rapidity-gap events which are
experimentally relevant at HERA.
 
\section*{Acknowledgements}
We are particularly grateful to all the participants for the lively discussions
which lasted much later than foreseen, and to the organizers who patiently
waited till the end of the session. Al Mueller was of precious help for
mastering beautifully the discussion and, in particular for asking the first
questions to the brave experimentalists present in the audience. Thanks to all!

One of us (R.P.) signals that this work was supported in part by the EU Fourth
 Framework Programme `Training and Mobility of Researchers', 
Network `Quantum Chromodynamics
and the Deep Structure of Elementary Particles', 
contract FMRX-CT98-0194 (DG 12 - MIHT).
Another one's (B.Z.K.) work was partially supported by European Network:
Hadronic Physics with Electromagnetic Probes, No FMRX CT96-0008,
and by INTAS grant No 93-0239.

\end{document}